\documentclass[global]{svjour}
\usepackage[utf8]{inputenc}

\usepackage{epsfig}
\usepackage{amsmath,amssymb,amsfonts}   
\numberwithin{equation}{section}
\usepackage{color}
\usepackage{graphicx,bm}
\usepackage{enumitem}

\renewcommand{\u}{{\mathbf u}}
\newcommand{\U}{{\mathbf U}}

\newcommand{\n}{{\mathbf n}}
\newcommand{\calN}{{\mathcal N}}
\newcommand{\calA}{{\mathcal A}}

\usepackage[numbers]{natbib}
\journalname{Bulletin of Mathematical Biology}

\begin{document}

\author{Hyunjoong Kim \inst{1} and Paul C. Bressloff \inst{1}}

\institute{Department of Mathematics, University of Utah, Salt Lake
  City, UT 84112 USA}
\titlerunning{Stochastic Turing pattern formation}
\title{Stochastic Turing pattern formation in a model with active and passive transport}

\date{\today}

\maketitle
\begin{abstract}
We investigate Turing pattern formation in a stochastic and spatially discretized version of a reaction diffusion advection (RDA) equation, which was previously introduced to model synaptogenesis in \textit{C. elegans}. The model describes the interactions between a passively diffusing molecular species and an advecting species that switches between anterograde and retrograde motor-driven transport (bidirectional transport). Within the context of synaptogenesis, the diffusing molecules can be identified with the protein kinase CaMKII and the advecting molecules as glutamate receptors. The stochastic dynamics evolves according to an RDA master equation, in which advection and diffusion are both modeled as hopping reactions along a one-dimensional array of chemical compartments. Carrying out a linear noise approximation of the RDA master equation leads to an effective Langevin equation, whose power spectrum provides a means of extending the definition of a Turing instability to stochastic systems, namely, in terms of the existence of a peak in the power spectrum at a non-zero spatial frequency. We thus show how noise can significantly extend the range over which spontaneous patterns occur, which is consistent with previous studies of RD systems.

\end{abstract}

\newpage
\section{Introduction} %-------------------------------------------------------------------------------------------

One major mechanism for self-organization within cells and between cells is the interplay between diffusion and nonlinear chemical reactions. Historically speaking, the idea that a reaction-diffusion (RD) system can spontaneously generate spatiotemporal patterns was first introduced by Turing in his seminal 1952 paper \cite{Turing52}. Turing considered the general problem of how organisms develop their structures during the growth from embryos to adults. He established the principle that two nonlinearly interacting chemical species differing significantly in their rates of diffusion can amplify spatially periodic fluctuations in their concentrations, resulting in the formation of a stable periodic pattern. The Turing mechanism for morphogenesis was subsequently refined by Gierer and Meinhardt \cite{Gierer72}, who showed that one way to generate a Turing instability is to have an antagonistic pair of molecular species known as an activator-inhibitor system, which consists of a slowly diffusing chemical activator and a quickly diffusing chemical inhibitor. Over the years, the range of models and applications of the Turing mechanism has expanded dramatically \cite{Murray08,Cross09,Walgraef97}.

Motivated by experimental studies of synaptogenesis in \textit{Caenorhabditis elegans} \cite{Rongo99,Hoerndli13,Hoerndli15}, we recently introduced a reaction-diffusion-advection (RDA) model for spontaneous pattern formation, which involved the interaction between a passively diffusing species and an advecting species that switches between anterograde and retrograde motor-driven transport (bidirectional transport). We identified the former species as the protein kinase CaMKII and the latter as the glutamate receptor GLR-1. Using linear stability analysis, we derived conditions on the associated nonlinear reaction functions for which a Turing instability can occur.  In particular, we showed that the dimensionless quantity $\gamma=\alpha D/{v^2}$ had to be sufficiently small for patterns to emerge, where $\alpha$ is the switching rate between motor states, $v$ is the motor speed, and $D$ is the diffusion coefficient of CaMKII. We thus established that patterns cannot occur in the fast switching regime ($\alpha \rightarrow \infty$), which is the parameter regime where the model effectively reduces to a two-component reaction-diffusion system. (Deterministic Turing pattern formation based on advecting species has also been considered within the context of chemotaxis \cite{Hillen96}. However, the model reduces to a traditional RD model in the fast switching limit.)
Numerical simulations of the model using experimentally-based parameters generated patterns with a wavelength consistent with the synaptic spacing found in {\em C. elegans}, after identifying the in-phase CaMKII/GLR-1 concentration peaks as sites of new synapses. Extending the model to the case of  a slowly growing 1D compartment, we subsequently showed how the synaptic density can be maintained during {\em C. elegans} growth, due to the insertion of new concentration peaks as the length of the compartment increases \cite{Brooks17}.

In this paper, we investigate how molecular (intrinsic) noise due to low copy numbers affects the RDA model. We proceed in an analogous fashion to previous studies of spontaneous pattern formation in stochastic RD systems \cite{Biancalani10,Butler09,Butler11,Woolley11,Schumacher13,McKane14,Biancalani17}. The latter incorporate diffusion into a stochastic biochemical network by discretizing space and treating diffusion as a set of hopping reactions. The resulting stochastic dynamics can then be represented in terms of a generalized RD master equation. Motivated by \cite{Lugo08,DeAnna10}, we carry out a linear noise approximation of the master equation, which leads to an effective Langevin equation. Its power spectrum provides a means of extending the definition of a Turing instability to stochastic systems, namely, in terms of the existence of a peak in the power spectrum at a non-zero spatial frequency. One thus finds that noise can significantly extend the range over which spontaneous patterns occur. That is, noise can increase the robustness of patterns in RD systems. This phenomenon has also been investigated experimentally in example biological systems \cite{Patti18,Karig18}. We will show that a similar result holds for the RDA system.

The structure of the paper is as follows. 
In section \ref{sect2}, we introduce the individual-based (or microscopic) model, which is a stochastic and spatially discretized version of the RDA model \cite{Brooks16}. Its stochastic dynamics evolves according to a chemical master equation, in which advection and diffusion are both modeled as hopping reactions. In the thermodynamic limit we recover a spatially discrete version of the deterministic (macroscopic) RDA model.
In section \ref{sect3}, we use linear stability analysis to derive conditions for a Turing instability in the deterministic model, and show how the results of \cite{Brooks16} are recovered in the continuum limit.
In section \ref{sect4}, we carry out a system-size expansion of the RDA master equation to obtain a mesoscopic model evolving according to a chemical Langevin equation.  Using a linear noise approximation, we obtain the corresponding power spectrum and use this to derive conditions for the occurrence of stochastic Turing patterns. Finally, in section \ref{sect5}, we highlight the result of this work and future directions.

%------------------------------------------------------------------------------------------------------------------------------------------------------
\section{Microscopic synaptogenesis model with active and passive transport} \label{sect2}

\begin{figure}[t!]
\begin{center}
\includegraphics[width=\linewidth]{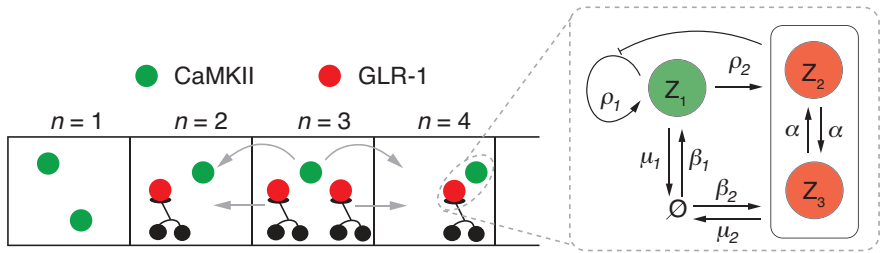}
\end{center}
\caption{
Illustration of the reactions of the microscopic synaptogenesis model with active and passive transport. CaMKII molecules ({\it green}) hop between neighboring compartments, whereas GLR-1 molecules ({\it red}) switch between left-moving and right-moving states.
CaMKll and GLR-1 molecules react according to a Gierer-Meinhardt reaction scheme\cite{Gierer72}.
}
\label{fig1}
\end{figure}

Consider a one-dimensional discrete lattice of compartments labeled $n = 1,2,\cdots$, N, as depicted in Fig.~\ref{fig1}. This lattice represents a neurite in the ventral cord of {\it C. elegans} during synaptogenesis.
Let $Z_1$ denote molecules of CaMKII that hop between neighboring sites at a rate $\kappa_1$. Similarly, let $Z_2$ ($Z_3$) denote molecules of GLR-1 that hop to the right (left) at a rate $\kappa_2$. These hopping reactions are the spatially discrete versions of passive diffusion and active motor-driven transport, respectively.
Take $\calN_{m,n}$ to represent the local number of $Z_m$ molecules in the $n$-th compartment with $m=1,2,3$.
The transport reactions of the species are specified according to
\begin{subequations} \label{hopp}
\begin{equation} \label{diff}
	Z_{1,n} \overset{\kappa_1}{\rightarrow} Z_{1,n\pm 1}, 
\end{equation}
\begin{equation} \label{dir}
	Z_{2,n} \overset{\kappa_2}{\rightarrow} Z_{2,n+1}, \quad Z_{3,n} \overset{\kappa_2}{\rightarrow} Z_{3,n-1}. 
\end{equation}
\end{subequations}
The active transport species locally switch their direction according to the two-state Markov chain
\begin{subequations} \label{chem}
\begin{equation} \label{swch}
	Z_{2,n} \underset{\alpha}{\overset{\alpha}{\rightleftharpoons}} Z_{3,n},
\end{equation}
where $\alpha$ is the switching rate.
In addition to the transport reactions given above, the actively and passively transported species interact locally through chemical reactions based on the Gierer and Meinhardt model \cite{Gierer72}:
\begin{equation} \label{brth}
	\emptyset \overset{\beta_1}{\rightarrow} Z_{1,n}, \quad \emptyset \overset{\beta_2/2}{\rightarrow}  Z_{m,n}, 
\end{equation}
\begin{equation} \label{deth} 
	Z_{1,n} \overset{\mu_1}{\rightarrow} \emptyset, \quad Z_{m,n} \overset{\mu_2}{\rightarrow}\emptyset,
\end{equation}
\begin{equation} \label{auto}
	2Z_{1,n} \overset{\rho_1 / (u_{2,n}+u_{3,n})}{\rightarrow} 2Z_{1,n}+ Z_{1,n}, \quad 2Z_{1,n} \overset{\rho_2/2}{\rightarrow} 2Z_{1,n} + Z_{m,n},  
\end{equation}
\end{subequations}
where $m = 2,3$.
These reactions describe the creation of a new molecule, a molecule being degraded from the system, and the autocatalysis of a molecule, respectively.
In particular, the autocatalysis of $Z_1$ is inhibited by the concentration of active transport molecules. Finally, each compartment is assumed to be well mixed.

% figure 1 --------------------------------

A systematic method for constructing the chemical master equation of the above microscopic model, is to identify the stoichiometric coefficients and propensity functions of the individual single-step reactions that appear in the deterministic mass action kinetics. That is, let $u_{m,n}=\calN_{m,n}/\Omega$, where $\Omega$ is the volume of each compartment, and consider the thermodynamic limit $\Omega\rightarrow \infty$ such that $u_{m,n}$ is finite. Suppose that
we index molecule species $Z_m$ in compartment $n$ by $I = 3(n-1)+m$ and label a single-step reaction by $\mu$.
We also introduce the integers $r_{I\mu}$ and $p_{I\mu}$ which describe, respectively, the number of reactants and products involved in reaction $\mu$.
The reaction $\mu$ can then be written in the general form
\begin{equation} \label{micro}
	\sum_I r_{I\mu} Z_I \overset{f_\mu}{\longrightarrow} \sum_I p_{I\mu} Z_I,
\end{equation}
where $f_\mu$ is the corresponding reaction rate.
The associated stoichiometric matrix element for species $I$ and reaction $\mu$ is defined by
\[
	S_{I\mu} = p_{I\mu} - r_{I\mu}.
\]
The mass action kinetic equations then take the compact form 
\begin{equation}
\label{rate}
\frac{du_I}{dt} =  \sum_{\mu} S_{I\mu}f_\mu(\mathbf{u}),
\end{equation}
with $[\mathbf{u}]_I = u_I$. Given the kinetic equations (\ref{rate}), the corresponding chemical master equation for the probability distribution
\[P(\n,t)={\mathbb P}[\calN_{I}(t)=n_I,I=1,\ldots,3N|\n(0)=\n_0]\]
takes the form
\begin{equation}
\label{master}
\frac{dP(\n,t)}{dt}=\Omega \sum_{\mu=1}^R\left (\prod_{I=1}^{3N}{\mathbb E}^{-S_{I\mu}}-1\right )f_\mu(\n/\Omega)P(\n,t),
\end{equation}
with $[\n]_I=\calN_I$ and $R$ is the total number of reactions.
Here ${\mathbb E}^{-S_{I\mu}}$ is a step or ladder operator such that for any function $g(\n)$,
\begin{equation}
{\mathbb E}^{-S_{I\mu}}g(\calN_1,\ldots,\calN_i,\ldots,\calN_{3N})=g(\calN_1,\ldots, {\mathcal N}_I-S_{I\mu},\ldots,{\mathcal N}_{3N}).
\end{equation}

We now determine the exact form of the stoichiometry matrix and the reaction vector by considering separately the transport hopping reactions and the local chemical reactions satisfying \eqref{hopp} and \eqref{chem}, respectively.
We label the eleven local chemical reactions listed in equation \eqref{chem} by the index $m_\text{loc} = 1,2,\cdots,11$.
We then label the local chemical reactions in compartment $n_\text{loc}$ by $\mu_1 = 11(n_\text{loc}-1)+m_\text{loc}$. 
Thus the corresponding stoichiometric matrix becomes
\begin{equation}
	S_{I\mu_1} ^\text{loc} = \delta_{n, n_\text{loc}} [\mathbf{L}]_{m,m_\text{loc}},
\end{equation}
where $\delta$ is the Kronecker's delta function and
\[
	\mathbf{L} = \left[\begin{array}{c c | c c c| c c c| c c c}
	0&0& 1&0&0& -1&0&0& 1&0&0 \\
	-1&1& 0&1&0& 0&-1&0& 0&1&0 \\
	1&-1& 0&0&1& 0&0&-1& 0&0&1
	\end{array}\right].
\]
Defining the local density vector $\mathbf{u}_n = [u_{1,n}, u_{2,n}, u_{3,n}]^T$, the reaction rate vector satisfies
\begin{equation}
	f_{\mu_1}^\text{loc} = [\mathbf{F}(\mathbf{u}_{n_\text{loc}})]_{m_\text{loc}}
\end{equation}
where
\[
	\mathbf{F}(\mathbf{u}) = \left[ \begin{array}{c c| c c c| c c c| c c c}
	\alpha u_2&\alpha u_3& \beta_1&\frac{\beta_2}{2}&\frac{\beta_2}{2}& \mu_1u_1&\mu_2u_2&\mu_2u_3& \frac{\rho_1u_1^2}{u_2+u_3}&
\frac{\rho_2}{2}u_1^2	& \frac{\rho_2}{2}u_1^2\end{array} \right]^T.
\]
%Here $\delta$ is the Kronecker's delta function.

In the same fashion, we determine the quantities for the hopping reactions. 
Labeling the hopping reactions in compartment $n_\text{hop}$ by $\mu_2 = 4(n_\text{hop} - 1) + m_\text{hop}$, where $m_\text{hop} = 1,\cdots,4$ represents \eqref{hopp} in the same order. Therefore, we have
\begin{equation}
	S_{I\mu_2}^\text{hop} = -\sum_{j = -1}^1 (-1)^j\delta_{n,n_\text{hop}+j} [\mathbf{H}_j]_{m,m_\text{hop}},
\end{equation}
where
\[
	\mathbf{H}_{+} = \left[\begin{array}{c c c c}
	1&0&0&0\\
	0&0&1&0\\
	0&0&0&0
	\end{array}\right], \quad \mathbf{H}_-  = \left[\begin{array}{c c c c}
	0&1&0&0\\
	0&0&0&0\\
	0&0&0&1
	\end{array}\right],
\]
and $\mathbf{H}_0 = \mathbf{H}_{+} + \mathbf{H}_{-}$, which reaction rates satisfies
\begin{equation}
	f_{\mu_2}^\text{hop} =\mathbf{K} \mathbf{u}_{n_\text{hop}}, \quad  \mathbf{K} = \left[\begin{array}{c c c}
	\kappa_1& 0&0\\
	\kappa_1& 0&0\\
	0& \kappa_2&0\\
	0&0& \kappa_2
	\end{array}\right].
\end{equation}
Combining our results we can rewrite equation (\ref{rate}) as
\begin{equation}
\label{rate2}
\frac{du_I}{dt} =   \sum_{\mu_1} S_{I\mu_1}^\text{hop} f_{\mu_1}^\text{hop}(\mathbf{u}) + \sum_{\mu_2} S_{I\mu_2}^\text{loc} f_{\mu_2}^\text{loc}(\mathbf{u}).
\end{equation}
In particular, the local dynamics in compartment $n$ takes the following explicit form:
\begin{equation} \label{macr}
	\frac{d\mathbf{u}_n}{dt} =\calA_n(\u):= \mathbf{H}_+\mathbf{K}(\mathbf{u}_{n-1} - \mathbf{u}_n) + \mathbf{H}_-\mathbf{K}(\mathbf{u}_{n+1} - \mathbf{u}_n) + \mathbf{G}(\mathbf{u}_n).
\end{equation}
Here the local reaction term is
\[
	\mathbf{G}(\mathbf{u}) \equiv \mathbf{L}\mathbf{F}(\mathbf{u}) = \left[\begin{array}{c}
	g_1(u_1,u_2,u_3)\\ g_2(u_1,u_2) -\alpha u_2 + \alpha  u_3\\ g_2(u_1,u_3) + \alpha u_2 - \alpha u_3 
	\end{array} \right].
\]
with the expressions for $g_1$ and $g_2$ given in appendix \ref{apdx:reaxn}. Equation (\ref{macr}) is the spatially discrete version of the RDA model analyzed in \cite{Brooks16,Brooks17}.

\subsection{Parameter values} \label{sect:pv}
The various model parameters can be obtained from the biological literature and previous modeling studies \cite{Brooks16, Brooks17}.
First, we set the compartment size to be $10 \mu m$, which is consistent with the relevant length scales of the ventral cord in {\em C. elegans} \cite{Rongo99}. The diffusion coefficient for CaMKII is around $D=0.01\ \mu$m$^2$/s. 
The average velocity of GRL-1 undergoing active transport along the ventral cord is 1-2 $ \mu$m/s, the average step size of the kinesin-3 family of motors is $0.01\ \mu$m, and the run length is typically 5-10 $\mu$m \cite{Monteiro12, Hoerndli13,Arpag19}. From this, we infer that the passive and active hopping rates are $\kappa_1\sim \kappa_2 \sim 100$/s, the switching rate is $\alpha \sim$ 0.1-0.5/s, and $N$ = 1000.
Additionally, note that the conditions for stability in the homogeneous steady state are satisfied in this model because the turnover rate of GLR-1 is approximately four times that of CaMKII \cite{Hanus13, Brooks16}. Hence, we also take $\mu_1 = 0.25$/s and $\mu_2 = 1$/s.

\section{Deterministic pattern formation in the macroscopic model} \label{sect3}

In this section we use linear stability analysis to derive conditions for a Turing instability in the deterministic model given by equation (\ref{macr}), and show how we recover the results of \cite{Brooks16} in the continuum limit. This will then be used as a baseline to investigate the effects of noise in section 4.
\subsection{Linear stability analysis and dispersion curves}

Setting $\mathbf{u}_n(t) = \mathbf{u}^*$ in equation (\ref{macr}), the spatially homogeneous steady-state solution $\mathbf{u}^*$ satisfies
\begin{equation} \label{req}
	0 = \mathbf{G}(\mathbf{u}^*),
\end{equation}
which becomes
\begin{equation}
	(\beta_1 - \mu_1 u_1^*)(\beta_2+\rho_2u_1^{*2}) + \mu_2\rho_1 u_1^{*2} = 0, \quad u_2^* = u_3^* = \frac{\beta_2 + \rho_2 u_1^{*2}}{2\mu_2}.
\end{equation}
Note that the cubic equation for $u_1^*$ always has a positive real root and thus $u_2^*$ and $u_3^*$ are also positive. Therefore, there is at least one spatially homogeneous steady-state solution.
We now linearize about the fixed point $\mathbf{u}^*$ by setting
\[
	\mathbf{u}_n(t) = \mathbf{u}^* + \mathbf{w}_n(t),
\]
which gives the linear equation
\begin{equation} \label{linz}
	\frac{d\mathbf{w}_n}{dt} = \mathbf{H}_+\mathbf{K}(\mathbf{w}_{n-1} - \mathbf{w}_n) + \mathbf{H}_-\mathbf{K}(\mathbf{w}_{n+1} - \mathbf{w}_n) + \nabla\mathbf{G}(\mathbf{u}^*)\mathbf{w}_n.
\end{equation}
Here the linearized local chemical reaction matrix takes the form of
\[
	\nabla \mathbf{G}(\mathbf{u}^*) = \left[\begin{array}{c c c}
	g_{1,1} & g_{1,2} & g_{1,2} \\
	g_{2,1} & g_{2,2} -\alpha & \alpha \\
	g_{2,1} & \alpha & g_{2,2} - \alpha
	\end{array}\right],
\]
where $g_{j,m}$ describes the partial derivative of $h_j$ with respect to $u_m$ at the homogeneous solution; see  appendix \ref{apdx:reaxn} for the exact form of the derivatives.
We have used the fact that $g_{1,2} = g_{1,3}$, $g_{2,1} = g_{3,1}$, and $g_{2,2} = g_{3,3}$ at the fixed point.
In the absence of spatial terms, the linearized system \eqref{linz} reads
\begin{equation}
	\frac{d\mathbf{w}}{dt} = \nabla\mathbf{G}(\mathbf{u}^*) \mathbf{w},
\end{equation}
where $\mathbf{w}$ has solutions of the form $\mathbf{w} \propto e^{\lambda t}$.
The eigenvalues $\lambda$ of this system satisfy the characteristic equation $\mbox{det}[\nabla\mathbf{G}(\mathbf{u}^*)-\lambda {\bf I}]=0$.
This has solutions
\begin{equation}
	\lambda = g_{2,2} - 2\alpha,	
\end{equation}
and
\begin{equation}
	\lambda = \frac{1}{2}\left(g_{1,1}+g_{2,2} \pm \sqrt{(g_{1,1}+g_{2,2})^2 -8(g_{1,1}g_{2,2}-g_{1,2}g_{2,1})}\right)
\end{equation}
We require the steady state to be stable in the absence of spatial components, which means $\text{Re}(\lambda) < 0$. Since $g_{2,2} = - \mu_2 < 0$, the conditions for $\lambda$ become
\begin{equation} \label{stabineq}
	g_{1,1} + g_{2,2} <0, \quad g_{1,1}g_{2,2} - g_{1,2}g_{2,1}>0.
\end{equation}

Now we consider the stability of the full system with respect to spatially periodic perturbations by setting
\[
	\mathbf{u}_n(t) = \mathbf{u}^* + \mathbf{w}(k) e^{\lambda t} e^{ikn},
\]
where $k = 2\pi k_0/N$, $k_0 = 0,1,\cdots, N-1$. We have imposed periodic boundary conditions on the lattice. This gives the matrix equation
\begin{equation}
	\lambda \mathbf{w}(k) = \mathcal{L}(k) \mathbf{w}(k),
\end{equation}
where the linear operator takes the form
\begin{equation} \label{evlin}
	\mathcal{L}(k) = (e^{-ik} \mathbf{H}_+ + e^{ik}\mathbf{H}_- - \mathbf{H}_0)\mathbf{K} + \nabla \mathbf{G}(\mathbf{u}^*).
\end{equation}
Thus $\lambda$ satisfies the characteristic equation
\begin{equation} \label{chareq}
	0 = \det[ \mathcal{L}(k) - \lambda \mathbf{I}] \equiv p(k,\lambda).
\end{equation}
Introducing 
\[
	q(k) = \cos(k) - 1,
\]
then the characteristic polynomial can be written as 
\begin{equation} \label{chareqEx}
	p(k,\lambda) = -\lambda^3 + p_2(q(k)) \lambda^2+  p_1(q(k)) \lambda+ p_0(q(k)),
\end{equation}
where the coefficient $p_j$ are polynomial functions of $q$:
\begin{equation} \label{pjq}
	p_j(q) = \sum_{m=0}^2 p_{j,m} q^m.
\end{equation}
The exact form of $p_{j,m}$ is given in appendix \ref{apdx:char}. Since $q(2\pi - k) = q(k)$, we see that the characteristic equation satisfies
\begin{equation}
	p(2\pi - k,\lambda) = p(k,\lambda).
\end{equation}
Moreover, the characteristic polynomial has three eigenvalues for a given $k$.
If there is an eigenvalue whose real part is positive, then the homogeneous solution is unstable with respect to spatial perturbations with wavenumber $k$, which generate a Turing instability.

\begin{figure}[t!]
\begin{center}
\includegraphics[width=\linewidth]{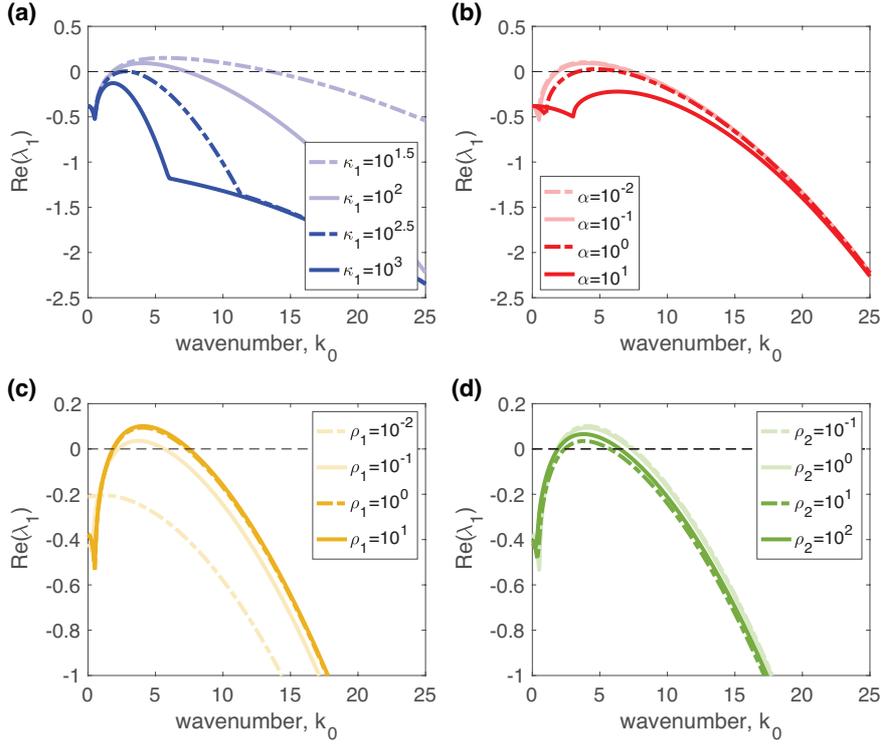}
\end{center}
\caption{Principal dispersion curves for the deterministic model. 
$\lambda_1$ represents the eigenvalue of the linearized system with the largest real part for a given wavenumber $k=2\pi k_0/N$, where $N = 1000$.
Dispersion curves are plotted as a function of $k_0$ for various parameter values: \textbf{(a)} passive transport rate $\kappa_1$; \textbf{(b)} switching rate $\alpha$; \textbf{(c)} autocatalysis rate for the passively transported particles $\rho_1$; and \textbf{(d)} autocatalysis rate for the actively transported particles $\rho_2$. Baseline parameters used in this and subsequent figures are as follows: $\alpha=0.1$/s, $\kappa_1 =\kappa_2= 100$/s, $\beta_1 = \beta_2 = 0.02$/s, and $\rho_1 = \rho_2 = 1$/s.
}
\label{fig2}
\end{figure}

In order to determine conditions for a Turing instability, we focus on the eigenvalue $\lambda_1(k)$ with the largest real part for a given $k$.
If there exists a range of $k$, $k>0$,  such that $\text{Re}(\lambda_1(k)) >0$, then the homogeneous solution supports a Turing instability.
The quantity $\text{Re}(\lambda_1(k))$ as a function of $k$ is usually referred as a dispersion curve for the system. 
The dispersion curves for various parameter values are depicted in Fig.~\ref{fig2}.
It can be seen that decreasing the passive transport rate $\kappa_1$ or the switching rate of bidirectional transport $\alpha$ leads to a Turing instability, see Fig.~\ref{fig2}(a) and (b), respectively. A Turing instability is also induced by increasing
$\rho_1$, while the dispersion curves are insensitive to changes in $\rho_2$, see Fig.~\ref{fig2}(c) and (d).
It can also be proven that the coefficients $p_{j,m}$ are independent of both $\rho_1$ and $\rho_2$ when $\beta_1 = \beta_2 = 0$, see appendix \ref{apdx:char}, and thus so are the dispersion curves.

% figure 2 --------------------------------

\begin{figure}[b!]
\begin{center}
\includegraphics[width=\linewidth]{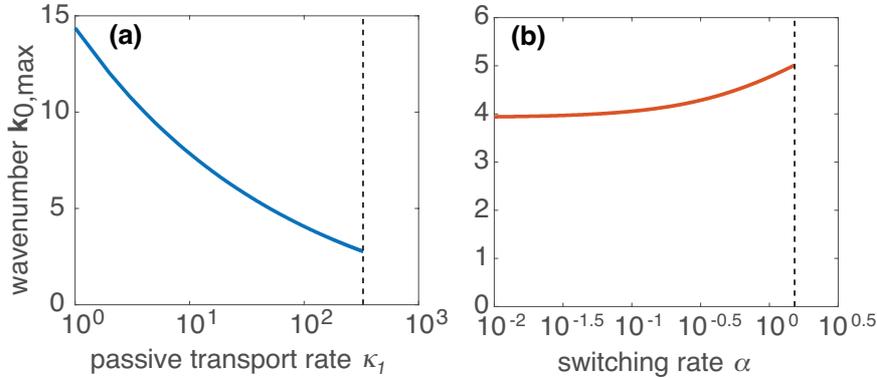}
\end{center}
\caption{Parameter dependence of $k_{\max}=2\pi k_{0,\max}/N$, where $k_{\max}$ is the wavenumber at the maximum of the principal dispersion curve $\text{Re}(\lambda_1(k))$, with $\text{Re}(\lambda_1(k_{\max})) \geq 0$.
Plot of $k_{0,\max}$ as a function of \textbf{(a)} the passive transport rate $\kappa_1$; and \textbf{(b)} the switching rate $\alpha$,. The dotted lines show the parameter value where $\text{Re}(\lambda_1)$ becomes negative for all $k_0$ (critical points). 
Other parameters as in Fig.~\ref{fig2}.
}
\label{fig3}
\end{figure}

We now derive criteria for the homogeneous solution to become unstable with respect to spatially periodic patterns.
In terms of the dispersion curve, we want to find conditions that ensure the curve touches zero at a single wavenumber in the range $k \in (0,\pi]$ (marginal stability).
One necessary condition is that there exists a unique wavenumber $k \in (0,\pi]$ for which there is a simple zero eigenvalue $\lambda=0$.
That is, there is a unique $q \in[-2,0]$ such that $p_0(q) = 0$.
Since $p_0(q)$ in equation \eqref{pjq} is a quadratic function, we obtain the condition
\begin{equation}
\label{critc}
	p_{0,1}^2 - 4 p_{0,2} p_{0,0} = 0,
\end{equation}
supplemented by the constraint
\begin{equation}
	-2 \leq - \frac{p_{0,1}}{2p_{0,2}} \leq 0.
\end{equation}
The set of parameters satisfying \eqref{critc} is often referred to as a critical curve, and crossing this critical curve signals the onset of a Turing instability. 
Above the critical curve there exists a band of unstable modes, which is dominated by the fastest growing mode. The corresponding dominant wavenumber $k_{\max}$ is at the maximum of the corresponding dispersion curve $\text{Re} (\lambda_1(k))$. Note that $k_{\max}$ is a continuous function of model parameters such as $\kappa_1$ and $\alpha$, as illustrated in Fig.~\ref{fig3}. 

In general, the characteristic equation \eqref{chareqEx} has one real root and two complex conjugate roots. However, in order to eliminate the case of a Turing-Hopf bifurcation, we must check that a pair of imaginary roots $\pm i\psi$ cannot occur at some value of $q \in [-2, 0]$. Setting $\lambda(q) = i\psi(q)$ in \eqref{chareqEx} for real $\psi(q)$ and equating real and imaginary parts yields the equations
\[
	p_0(q) - p_2(q)\psi^2(q) = 0, \quad p_1(q) + \psi^2(q) = 0.
\]
This can be deduced to the following cubic equation
\begin{equation} \label{THb}
	p_\text{TH}(q) \equiv p_0(q) + p_1(q)p_2(q) = 0,
\end{equation}
which can be written as
\[
	p_\text{TH}(q) = \sum_{m=0}^3 p_{\text{TH},m} q^m.
\]
Using the stability condition of the linearized system without spatial terms \eqref{stabineq}, one can show that
\begin{equation} \label{pth1}
	p_{\text{TH},3} < 0, \quad p_{\text{TH},2} > 0, \quad p_{\text{TH},1} < 0,
\end{equation}
and
\begin{equation} \label{pth2}
	p_\text{TH}(0) =p_{\text{TH},0} >0,
\end{equation}
a proof of which can be found in appendix \ref{apdx:char}.
Taking derivatives yields the quadratic function
\[
	p_\text{TH}'(q) = 3 p_{\text{TH},3} q^2 + 2 p_{\text{TH},2} q + p_{\text{TH},1},
\]
which satisfies
\begin{equation}
	p_\text{TH}'(q) <0, \quad q \in [-2,0],
\end{equation}
according to \eqref{pth1}. That is, $p_\text{TH}(q)$ is monotonically decreasing in $[-2,0]$. Therefore, \eqref{pth2} follows that
\begin{equation}
	p_\text{TH}(q) >0 ,\quad q \in [-2,0],
\end{equation}
which implies that a pair of complex conjugate roots cannot cross the imaginary axis.

\subsection{Relation to the continuum model}

Since our macroscopic transport model is the spatially discretized version of the RDA model of \cite{Brooks16}, the system of equations \eqref{macr} recovers the transport equations of \cite{Brooks16} in the continuum limit. Let the length of each compartment be $L/N$ and
set $\Delta x = L/N$. Taking $\Delta x\rightarrow 0$ and $N\rightarrow \infty$ for fixed $L$, with
\begin{equation}
\label{kap}
	\kappa_1 = \frac{D}{\Delta x^2}, \quad \kappa_2 = \frac{v}{\Delta x},
\end{equation}
one finds that $\mathbf{u}_n(t)$ converges to $\mathbf{u}(x,t)$ such that
\begin{equation}
	\frac{\partial \mathbf{u}}{\partial t} = \begin{bmatrix}
		D&0&0\\0&0&0\\0&0&0
	\end{bmatrix} \frac{\partial^2 \mathbf{u}}{\partial x^2} + \begin{bmatrix}
		0&0&0\\0&-v&0\\0&0&v
	\end{bmatrix}\frac{\partial \mathbf{u}}{\partial x} + \begin{bmatrix}
		0&0&0\\ 0&-\alpha&\alpha\\ 0&\alpha&-\alpha
	\end{bmatrix} \mathbf{u} + \mathbf{G}(\mathbf{u}).
\end{equation}
Moreover, the linear operator \eqref{evlin} for the eigenvalue problem also converges to the one in \cite{Brooks16}. Setting $k = k_c \Delta x$, the spatial term becomes
\begin{equation}
	(e^{-ik} \mathbf{H}_+ + e^{ik}\mathbf{H}_- - \mathbf{H}_0)\mathbf{K} \to \begin{bmatrix}
		-k_c^2 D&0&0\\ 0&-ik_c v&0\\ 0&0&ik_cv
	\end{bmatrix}
\end{equation}
as $\Delta x \to 0$, which proves our statement.

%------------------------------------------------------------------------------------------------------------------------------------------------------
\section{Stochastic pattern formation in the mesoscopic model} \label{sect4} 

Unfortunately, it is not possible to analyze the RDA master equation (\ref{master}) directly. However, as we show in this section, we can use it to explore the effects of molecular noise on spontaneous pattern formation by carrying out a system-size expansion along analogous lines to previous studies of RD master equations \cite{Biancalani10,Butler09,McKane14}. This will generate a corresponding mesoscopic model that evolves according to a chemical Langevin equation. We can then use spectral theory to derive condition for stochastic Turing patterns.

\subsection{System-size expansion}

The basic idea of the system-size expansion is to set $f_{\mu}(\n/\Omega)P(\n,t)\rightarrow f_{\mu}(\u)p(\u,t)$ with $\u=\n/\Omega$ treated as a continuous vector so that for any smooth function $h(\u)$,
\begin{eqnarray*}
 \prod_{I=1}^{3N}{\mathbb E}^{-S_{I\mu}}h(\u)&=&h(\u-{\bf S}_{\mu}/\Omega)\\
 &=&h(\u)-\Omega^{-1}\sum_{I=1}^{3N}S_{I\mu}\frac{\partial h}{\partial u_I}+\frac{1}{2\Omega^2}\sum_{I,J=1}^{3N}S_{I\mu}S_{J\mu}\frac{\partial^2h(\u)}{\partial u_I\partial u_J}\nonumber \\
 &&\qquad 
+O(\Omega^{-3}).
\end{eqnarray*}
Carrying out a Taylor expansion of the master equation to second order thus yields a multivariate Fokker-Planck equation of the Ito form:
\begin{equation}
\label{FPg}
\frac{\partial p}{\partial t}=-\sum_{I=1}^{3N}\frac{\partial A_I(\u)p(\u,t)}{\partial u_I}+\frac{1}{2\Omega}\sum_{I,J=1}^{3N}\frac{\partial^2 C_{IJ}(\u)p(\u,t)}{\partial u_I\partial u_J},
\end{equation}
where
\begin{equation}
\label{VBg}
A_I(\u)=\sum_{\mu=1}^R S_{I\mu}f_{\mu}(\u),\quad C_{IJ}(\u)=\sum_{\mu=1}^R S_{I\mu}S_{J\mu}f_{\mu}(\u).
\end{equation}
The FP equation (\ref{FPg}) corresponds to the Ito SDE
\begin{equation}
\label{Langg}
dU_I=A_I(\U)dt+\frac{1}{\sqrt{\Omega}}\sum_{\mu=1}^R K_{I\mu}(\U)dW_{\mu}(t),
\end{equation}
where $W_{\mu}(t)$ are independent Wiener processes
\begin{equation}
\langle dW_{\mu}(t)\rangle = 0,\quad \langle dW_{\mu}(t)dW_{\nu}(t')\rangle =\delta_{\mu,\nu}\delta(t-t')dt \, dt',\end{equation}
 and ${\bf C}={\bf K}{\bf K}^T$, that is,
\begin{equation}
K_{I\mu}=S_{I\mu}\sqrt{f_{\mu}(\u)}.
\end{equation}

\begin{figure}[t!]
\begin{center}
\includegraphics[width=\linewidth]{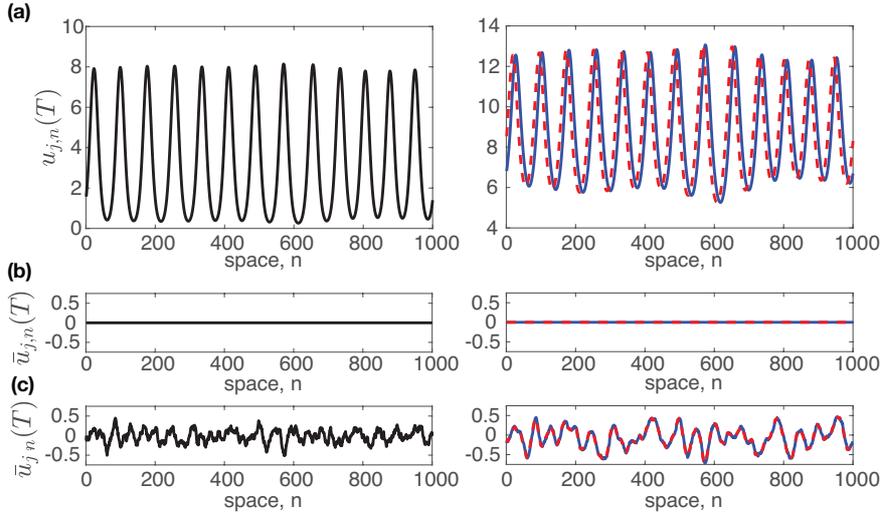}
\end{center}
\caption{
Pattern formation in the synaptogenesis model of the passively diffusing species CaMKII (\textit{left}) and the actively transported species GLR-1 (\textit{right}).
\textbf{(a)} Macroscopic spatial profiles of the deterministic concentrations evolving according to equation \eqref{macr}. Parameters are chosen so that the system operates in a regime predicted to exhibit a Turing pattern according to Fig.~\ref{fig2}: $\kappa_1 = 10^1$/s, $\alpha = 10^1$/s, and $T = 500$s.
\textbf{(b)} Corresponding deterministic plots for $\alpha =10^2$/s so that the system operates outside the regime for Turing patterns.
\textbf{(c)} Mesoscopic profiles of the SDE \eqref{Langg} obtained by time-averaging the concentration with normalization $\overline{u}_{j,n}(T) = \Omega^{1/2}\int_0^T u_{j,n}(t) - u_j^*dt$. Parameters are the same as (b).
}
\label{fig4}
\end{figure}

\begin{figure}[t!]
\begin{center}
\includegraphics[width=\linewidth]{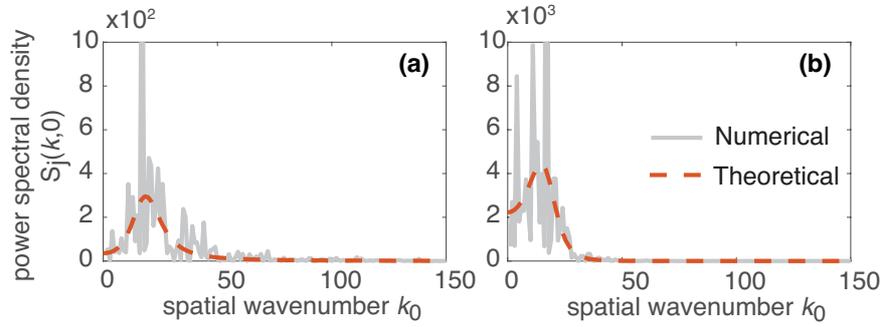}
\end{center}
\caption{
Plot of power spectral densities of \textbf{(a)} passively diffusing species $j = \text{diff}$ and \textbf{(b)} actively transporting species $j = \text{adv}$. The \textit{gray curves} depicts the densities computed by numerical results in Fig. \ref{fig4}(c). The \textit{orange-dotted curves} represents the theoretical results from \eqref{psd0} and \eqref{psd00}. Parameters for the theory are chosen to be the same as Fig. \ref{fig4}(c).}
\label{fig5}
\end{figure}

It will be convenient to formally rewrite the SDE as the chemical Langevin equation 
\begin{equation}
	\frac{du_I}{dt} = \sum_{\mu} S_{I\mu} f_\mu(\mathbf{u}) + \frac{1}{\sqrt{\Omega}} \eta_I (\mathbf{u},t).
\end{equation}
Here the Gaussian white noise terms $\eta_I$ have zero mean and correlations
\[
	\langle \eta_{I_1}(\mathbf{u},t_1) \eta_{I_2}(\mathbf{u},t_2)\rangle = \delta(t_1 - t_2) C_{I_1 I_2}(\mathbf{u}),
\] 
The corresponding Langevin equation for the local densities in the $n$-th compartment is given by
\begin{equation}
\label{lan2}
	\frac{d\mathbf{u}_n}{dt} = \mathcal{A}_n(\mathbf{u}) + \frac{1}{\sqrt{\Omega}} \boldsymbol \eta_n(\mathbf{u},t),
\end{equation}
with $\mathcal{A}_n$ defined in equation (\ref{macr}) and $\boldsymbol \eta_n$ a 3-vector with zero mean and correlation matrix
\begin{equation} \label{mesocor}
	\langle \boldsymbol \eta_{n_1} (\mathbf{u},t_1) \boldsymbol \eta_{n_2}^T (\mathbf{u},t_2) \rangle= \delta (t_1 - t_2) \mathbf{C}_{n_1n_2}(\mathbf{u}),
\end{equation}
where the matrix $3\times 3$ matrix $\mathbf{C}_{n_1n_2}$ for fixed $n_1,n_2$ takes the form
\begin{align*}
	\mathbf{C}_{n_1n_2}(\mathbf{u}) &= \delta_{n_1,n_2}\left[\mathbf{B}_{1,n_1}(\mathbf{u}) + \mathbf{B}_{2,n_1}(\mathbf{u})\right] + \delta_{n_1-1,n_2} \mathbf{B}_{3,n_1-1}(\mathbf{u}) \\
	&\quad + \delta_{n_1+1,n_2} \mathbf{B}_{3,n_1}  (\mathbf{u}).
\end{align*}
The explicit form of the $3\times 3$ matrices on the right-hand side are given in appendix \ref{apdx:meso}.
Numerical simulations of the SDE (\ref{Langg}) for large $\Omega$ are shown in Fig.~\ref{fig4}. Time-averaged concentration profiles clearly indicate a spatial pattern  in the stochastic model, even though the system operates in a parameter regime where the deterministic model does not exhibit Turing patterns. In Fig. \ref{fig5} we show corresponding numerical plots of the power spectral densities, which have a peak at a non-zero wavenumber that is consistent with theoretical predictions based on a linear noise approximation, see below.

\subsection{Linear noise analysis and power spectrum}
The chemical Langevin equation (\ref{lan2}) reduces to the deterministic rate equation (\ref{macr}) in the limit $\Omega \rightarrow \infty$.
This implies that in the presence of Gaussian fluctuations (finite $\Omega$) the homogeneous solution of the deterministic system is still unstable to spatially varying perturbations above the critical curve of the deterministic Turing instability.
However, it is also possible that fluctuations induce pattern forming instabilities in the sub-critical region of the deterministic model.
In order to explore this issue, we linearize the SDE (\ref{lan2}) about the homogeneous solution and determine its power spectrum.

Linearizing about $ \mathbf{u}^* $ by setting
\begin{equation}
	\mathbf{u}_n(t) = \mathbf{u}^* + \frac{1}{\sqrt{\Omega}} \mathbf{v}_n(t),
\end{equation}
 yields the linear Langevin equation
\begin{equation} \label{lna-p}
	\frac{d\mathbf{v}_n}{dt} = \nabla \mathcal{A}_n(\mathbf{u}^*)\mathbf{v}_n + \boldsymbol \xi_n(t).
\end{equation}
Here the white noise satisfies the correlation matrix \eqref{mesocor} at the homogeneous solution $\mathbf{u}^*$. We now
introduce the following discrete Fourier transform with respect to space
\[
	\mathbf{V}(k,t) = \sum_l \mathbf{v}_l(t) e^{-ikl}, \quad \mathbf{v}_n(t) = \frac{1}{N} \sum_k \mathbf{V}(k,t) e^{ikn},
\]
where $k = 2\pi k_0/N$, $k_0 = 0,1,\cdots N-1$ (for periodic boundary conditions).
Taking the discrete Fourier transformation of equations \eqref{lna-p} and using the identity $\sum_n e^{-i(k+k')n} = N\delta_{k,-k'}$, gives
\begin{equation} \label{lna-t}
	\frac{d\mathbf{V}(k,t)}{dt} = \mathcal{L}(k) \mathbf{V}(k,t) + \boldsymbol\Xi(k,t),
\end{equation}
The zero mean correlation matrix satisfies
\begin{equation}
	\langle \boldsymbol\Xi(k,t) \boldsymbol\Xi^T(k',t')\rangle = N\delta(t-t') \delta_{k,-k'} \mathbf{D}(k), 
	\end{equation}
	with
	\begin{equation}  \mathbf{D}(k) = \mathbf{B}_1^* + \mathbf{B}_2^*+ 2 \cos(k) \mathbf{B}_3^*,
\end{equation} 
where $\mathbf{B}_l^* = \mathbf{B}_{l,n}(\mathbf{u}^*)$, $l=1,2,3$, which is independent of $n$ since $\u^*$ is the fixed point.
Next, taking the temporal Fourier transform and rearranging yields
\begin{equation}
	\boldsymbol \Phi (k,w) \widehat{\mathbf{V}}(k,w) = \widehat{\boldsymbol\Xi}(k,w), \quad \boldsymbol \Phi (k,w) = - iw \mathcal{I} - \mathcal{L}(k),
\end{equation}
where $\mathcal{I}$ is the identity matrix.
We thus obtain the correlation matrix
\begin{align}
	\langle \widehat{\mathbf{V}}(k,w) \widehat{\mathbf{V}}^T(k',w') \rangle &= \langle \boldsymbol \Phi^{-1} (k,w), \widehat{\boldsymbol\Xi}(k,w) \widehat{\boldsymbol\Xi}^T(k',w') [\boldsymbol \Phi^{-1} (k',w')]^T\rangle \nonumber\\
	&= \delta(w+w') \delta_{k,-k'} \boldsymbol \Psi(k,w),
\end{align}
where
\[
	\boldsymbol \Psi (k,w) = \boldsymbol \Phi^{-1} (k,w) \mathbf{D}(k) ( \boldsymbol \Phi^{\dagger})^{-1} (k,w).
\]
Defining the power spectral density of the diffusing species according to
\[
	\left\langle \widehat{\mathbf{V}}_1(k,w) \widehat{\mathbf{V}}_1(k',w') \right\rangle = \delta(w+w') \delta_{k,-k'} S_\text{diff}(k,w),
\]
we deduce that
\begin{equation} \label{psd0}
	S_\text{diff}(k,w) = \Psi_{11}(k,w).
\end{equation}
Similarly, the power spectral density of the advecting species is defined by
\[
	\left\langle \sum_{j=2,3}\widehat{\mathbf{V}}_j(k,w) \sum_{j=2,3}\widehat{\mathbf{V}}_j(k',w')\right\rangle = \delta(w+w') \delta_{k,-k'} S_\text{adv}(k,w),
\]
which implies that
\begin{equation} \label{psd00}
	S_\text{adv}(k,w) = \sum_{j=2,3} \sum_{j'=2,3} \Psi_{j,j'}(k,w).
\end{equation}
Note that $S_\text{adv}$ is real because $\boldsymbol \Psi = \boldsymbol \Psi^\dagger$.

\begin{figure}[t!]
\begin{center}
\includegraphics[width=12cm]{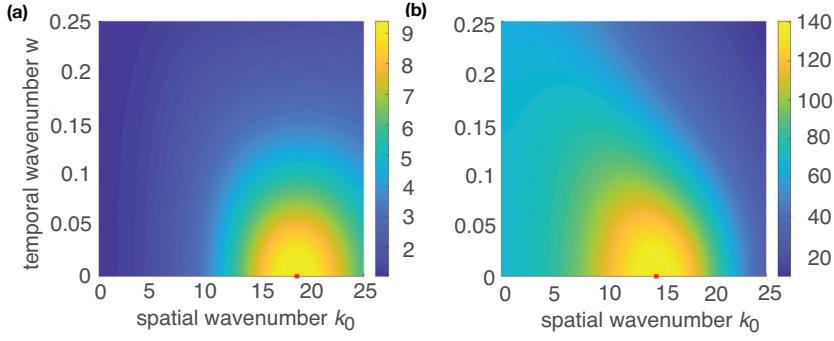}
\end{center}
\caption{Plot of the power spectral densities of the mesoscopic model as a function of spatial wavenumber $k = 2\pi k_0/N$ and temporal wavenumber $w$ for \textbf{(a)} diffusing species $S_\text{diff}(k,w)$ and \textbf{(b)} advecting species $S_\text{adv}(k,w)$. \textit{Red dots} represent the maximum of the power spectrum. Here we choose the same parameters as Fig.~\ref{fig4}(c).}
\label{fig6}
\end{figure}

Numerical plots of the theoretical power spectrum show that the power spectral densities are maximized at $k_0\neq 0$ and $w = 0$, as seen in Fig.~\ref{fig6}. Moreover, the power spectrum is decreasing with $w$ for given $k_0$ so that the fluctuations are characterized by spatial scale $k_0$. That is, there is a non-trivial wavenumber $k_{0,\max}$ maximizing the power spectral density at $w = 0$. This holds for various parameters, not just shown in specific parameter set in Fig.~\ref{fig6}. Another observation is that the power spectrum of the advecting species has a larger wavelength than the diffusing species, see Fig.~\ref{fig6}.

\subsection{Stochastic Turing pattern formation}
% figure 6 --------------------------------
\begin{figure}[t!]
\begin{center}
\includegraphics[width=\linewidth]{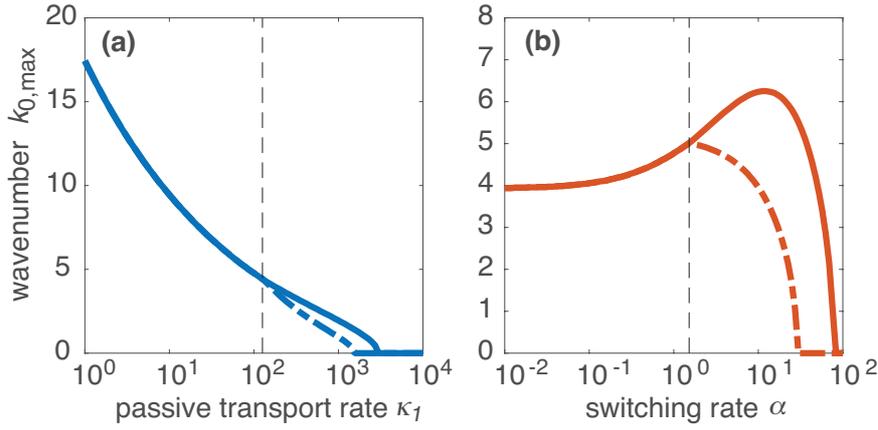}
\end{center}
\caption{Parameter dependence of $k_{\max}=2\pi k_{0,\max}/N$, where $k_{\max}$ is the wavenumber at the maximum of the power spectral density $S_j(k,0)$. Plots of $k_{0,\max}$ as a function of \textbf{(a)} the passive transport rate $\kappa_1$; and \textbf{(b)} the switching rate $\alpha$. The \textit{solid lines} and \textit{dotted lines} represents the case of $j= $ diff and $j= $ adv, respectively. (The curves merge for small $\kappa_1$ and $\alpha$.) The \textit{black dotted lines} show the parameter value where $\text{Re}(\lambda_1)$ becomes negative for all $k_0$ (critical points of deterministic model). The parameter values where $k_{0,\max}=0$ indicate the critical points for the stochastic model. Other parameters are as in Fig.~\ref{fig2}.}
\label{fig7}
\end{figure}

We can now construct a stochastic analog of the critical curve for a Turing instability in the deterministic system by investigating when the wavenumber $k_{0,\max}$ at the maximum of the power spectral density given by equation (\ref{psd00}) first vanishes, that is, $k_{0,\max} = 0$.
Numerical results in Fig.~\ref{fig7} show how $k_{0,\max}$ varies continuously with the model parameters $\kappa_1$ and $\alpha$. In each case, $k_{0,\max}$ becomes zero at a critical parameter value which we identify with the critical point at which stochastic Turing patterns disappear. In other words, the boundary of the parameter region for stochastic Turing pattern formation satisfies 
\begin{equation} \label{stcond}
	\left. \frac{d S_j(k,0)}{dk} \right|_{k=0} = 0,\quad \left. \frac{d^2 S_j(k,0)}{dk^2} \right|_{k=0} = 0.
\end{equation}
Consider the power spectral densities \eqref{psd0} and \eqref{psd00} with the following components,
\begin{equation} \label{psd1}
	\Psi_{jj'}(k,w) = D_{jj'}(k) \left|\frac{\det_{jj'}[\boldsymbol \Phi(k,w)]}{\det[\boldsymbol \Phi(k,w)]}\right|^2,\quad j,j' = 1,2,3,
\end{equation}
where $\det_{jj'}[A]$ represents the determinant of the submatrix of $A$ which does not include the $j'$th row and the $j$th column of $A$.
Substituting $q = \cos(k)-1$ into \eqref{psd1} and taking $w=0$, we define $W_j(q)$ such that
\[
	W_j(q (k)) = S_j(k,0).
\]
Using the chain rule we find that
\begin{align*}
	\left. \frac{d^2 S_j(k,0)}{dk^2} \right|_{k=0} &= \left[ W_j''(q) \left(\frac{dq}{dk}\right)^2 + W_j'(q) \frac{d^2q}{dk^2}\right]_{k=0}= -W_j'(0),
\end{align*}
which leads to the condition
\begin{equation} \label{stcondq}
	W_j'(0) = 0.
\end{equation}
In particular, direct computation shows that $W_1(q)$ takes the form of a rational function of $q$,
\begin{equation} \label{psdw}
	W_1(q) = [q\theta_{1,1} + \theta_{1,0}]\left[\frac{q\theta_{2,1} + \theta_{2,0}}{q^2\theta_{3,2} + q \theta_{3,1} + \theta_{3,0}}\right]^2.
\end{equation}
The explicit form of the coefficients are presented in appendix \ref{apdx:psd}.
Taking derivatives of $W_1$ at $q = 0$, we obtain the following condition for $k_{0,\max}=0$:
\begin{equation} \label{scritc}
	0 = \theta_{2,0} \left[2 \theta_{1,0} (\theta_{2,1}\theta_{3,0}-\theta_{2,0}\theta_{3,1})+ \theta_{1,1} \theta_{2,0}\theta_{3,0} \right].
\end{equation}

One interesting observation from Fig.~\ref{fig7} is that the curve of maximum $k_{0,\max}$ splits into two curves in the stochastic pattern forming region.
Note that the critical points of the deterministic model (depicted by the black dotted line in Fig.~\ref{fig7}) is determined by the following equation
\[
	\det[ \boldsymbol\Phi(k,0)] = 0,
\]
and the expression on the right-hand side is the common denominator term in equation \eqref{psd1}.
As the parameter $\kappa_1$ or $\alpha$ decreases, the denominator decays to zero and it dominates the behavior of the power spectrum. Therefore, the two curves in the stochastic model merge at the critical point of the deterministic model.
Furthermore, we note that the deterministic condition \eqref{chareq} for the critical wavenumber applies to all of the chemical species, whereas the stochastic version given by equations \eqref{psd0} and \eqref{psd00} represents the critical wavelength for each species separately. Indeed, 
Fig.~\ref{fig7} shows the existence of a parameter region where $k_{0,\max}$ is zero when $j = \text{adv}$ but not $j = \text{diff}$. 
This suggests that the passively diffusing species can make patterns even when the actively transported species does not.
However, we emphasize that the $j = \text{adv}$ case represents the sum of both directional transport subspecies, which can average out the patterning. It can be also confirmed by the fact that the real part of $\Psi_{2,3}$ does not need to be positive. 

\begin{figure}[t!]
\begin{center}
\includegraphics[width=.8\linewidth]{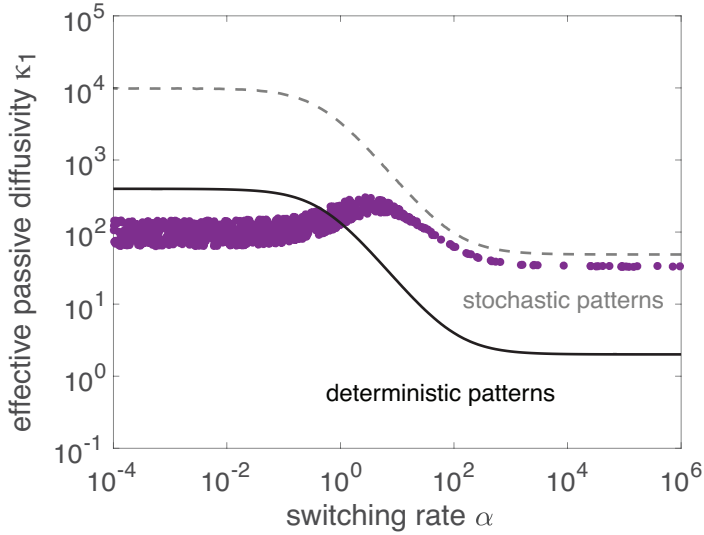}
\end{center}
\caption{
Deterministic and stochastic Turing instability regions in the $(\alpha,\kappa_1)$-plane. 
The {\it solid curve} satisfies the deterministic conditions \eqref{critc}, 
and the {\it dotted curve} satisfies the stochastic critical condition \eqref{scritc}.
{\it Purple dots} represent the parameter values for which $k_{0,\max} \in [3.5, 4.5]$. The latter range is consistent with the density of synaptic site found in experimental data \cite{Rongo99}.
Other parameter values are the same as Fig.~\ref{fig2}.
}
\label{fig8}
\end{figure}

The condition (\ref{scritc}) can now be used to determine bifurcation curves for the onset of stochastic Turing pattern formation. These can then be compared with the corresponding bifurcation curves of the deterministic model, which are determined by equations (\ref{critc}). An example stability diagram is shown in Fig. \ref{fig8}, which establishes that intrinsic noise can enlarge the parameter region over which Turing patterns occur.
Fig. \ref{fig8} shows that the deterministic region in the $(\alpha,\kappa_1)$-plane where Turing instabilities occur persists for all switching rates $\alpha$ provided that $\kappa_1$ is sufficiently small. It was previously shown that for the continuum RDA model with $\beta_1=\beta_2=0$, the dimensionless parameter $\gamma = \alpha D/v^2$ has to be sufficiently small for Turing patterns to occur \cite{Brooks16}. The analogous parameter in the deterministic compartmental model is $\gamma=\alpha \kappa_1/\kappa_2^2$, see equation (\ref{kap}). Suppose that we fix $\kappa_{1,2}$ and take the limit $\alpha\rightarrow \infty$. The deterministic condition \eqref{critc} then reduces to
\begin{equation} \label{ascond1}
	(2g_{2,2}\kappa_1 + g_{1,1} \kappa_2)^2 - 8\kappa_1\kappa_2(2g_{1,2}g_{2,1} - g_{1,1}g_{2,2}) = 0, 
	\end{equation}
	together with the constraint
	\begin{equation}0 \leq g_{1,1}\kappa_2 + 2 g_{2,2}\kappa_2 \leq 8\kappa_1 \kappa_2.
\end{equation}
Let $\kappa_{1,\det}^*$ be the solution of equation \eqref{ascond1}.
In particular, note that if $\beta_1 = \beta_2 = 0$, then
\[
	\kappa_{1,\det}^* = - \frac{\kappa_2 \mu_1}{2 \mu_2} <0,
\]
which means that horizontal asymptote of the deterministic stability curve is negative. That is, deterministic Turing patterns disappear in the limit $\alpha \rightarrow \infty$, as found previously \cite{Brooks16}.
Similarly, taking the limit $\alpha\rightarrow \infty$ in the stochastic Turing condition \eqref{scritc}, we again have a quadratic polynomial
\begin{equation} \label{ascondst}
	\zeta_2\kappa_1^2 + \zeta_1 \kappa_1 + \zeta_0 = 0,
\end{equation}
which determines the horizontal asymptote of the stochastic stability curve $\kappa_{1,\text{stoch}}^*$. The explicit form of the coefficients $\zeta_j, j=0,1,2,$ can be found in appendix \ref{apdx:psd}. For any $\beta_1$ and $\beta_2$, the sign of the coefficients satisfies
\[
	\zeta_2 = 2 g_{2,2}^2 u_1^* > 0, \quad \zeta_0 = - \mu_1 \rho_1\rho_2 \kappa_2 \frac{(1+u_1^*)u_1^{*3}}{(u_2^*+u_3^*)^2} <0,
\]
which means that there exists a positive $\kappa_{1,\text{stoch}}^*$. 
It follows that in the regime of fast switching, there are noise-induced patterns over the interval
\[\kappa_{1,\det}^*<\kappa_1<\kappa_{1,\text{stoch}}^*.\]

Finally, note that there exists a region in parameter space where the wavelength of the dominant pattern is consistent with the experimentally measured spacing of synapses in the ventral cord of {\em C. elegans}. This region is indicated by the purple dots in Fig. \ref{fig8}. The synaptic density is found to be around $3.7 \pm 0.1$ per $10 \ \mu$m \cite{Rongo99}, which corresponds to $k_{0,\max} \in [3.5, 4.5]$. It can be seen that intrinsic noise significantly enhances the region in which such patterns can be found (given that we are using log-log plots). Note that the baseline parameter values $\kappa_1=100/s$ and $\alpha=$ 0.1-0.5/s for {\em C. elegans} places the model within the purple band, but suggests that the system operates in a regime where the deterministic system also supports spatial patterns.  However, one important aspect of active transport that we have ignored in this paper is that the motor-GluR1 complexes can also stop moving for a few seconds before starting another run \cite{Hoerndli13}. This means that the effective mean-square displacement of the active particles is reduced. Since the effective diffusivity of the active particles scales as $D_{\rm eff}\sim v^2/\alpha$, it follows that the system could be pushed into the regime where only stochastic patterns occur. It should also be remembered that the hybrid transport model could have applications to other biological systems that operate in different parameter regimes. We hope to explore these issue further in future work.

\section{Discussion} \label{sect5} 

In this paper, we considered a stochastic and spatially discrete version of an RDA model that was originally introduced to model synaptogenesis in {\em C. elegans}. The latter is a hybrid reaction-transport model in which one chemical species passively diffuses while the other undergoes bidirectional active transport. Here we showed how intrinsic noise due to low copy numbers can enlarge the parameter region where Turing pattern formation occurs.  We proceeded in an analogous fashion to previous studies of RD systems, by constructing a chemical master equation in which the transport processes are represented as hopping reactions. Performing a system-size expansion of the master equation, we derived a chemical Langevin equation that approximated the effects of intrinsic noise in terms of Gaussian fluctuations about the corresponding deterministic model. Using a linear noise approximation, we calculated the resulting power spectral density and derived a condition for stochastic Turing pattern formation in terms of whether or not the density had a peak at a nonzero wavenumber. This allowed us to construct a stability diagram comparing the parameter regions that support deterministic and stochastic patterns, respectively. We also identified the region of parameter space that supports patterns whose wavelength are consistent with the spacing of synapses in the ventral cord of {\em C. elegans}. 

Although noise-induced pattern formation can broaden the parameter region over which a Turing pattern exists, the amplitude of the patterns are $O(\Omega^{-1/2})$, where $\Omega$ is the system size. Hence, the amplitude of a pattern in the fluctuation-driven regime is expected to be much smaller than in the deterministic regime, and thus might not be observable. However, in the case of RD systems, it has recently been shown how the giant amplification of fluctuation-driven patterns can occur in cases where there is an interplay between intrinsic noise and transient growth of perturbations about a spatially uniform state \cite{Biancalani17}. It would be interesting to explore an analogous amplification in RDA systems.

%------------------------------------------------------------------------------------------------------------------------------------------------------
%\section*{Appendices}
\appendix
\section{Exact forms of the macroscopic and mesoscopic models} 

\subsection{Reaction components of macroscopic model} \label{apdx:reaxn}
Multiplying $\mathbf{L}$ and $\mathbf{F}(\mathbf{u})$ in Sect. \ref{sect2} yields
\begin{equation}
	\mathbf{L} \mathbf{F}(\mathbf{u}) = \left[\begin{array}{c}
	\beta_1 - \mu_1 u_1 + \frac{\rho_1 u_1^2}{u_2+u_3} \\
	\frac{\beta_2}{2} - \mu_2 u_2 + \frac{\rho_2 u_1^2}{2} -\alpha u_2 + \alpha u_3\\
	\frac{\beta_2}{2} - \mu_2 u_3 + \frac{\rho_2 u_1^2}{2} +\alpha u_2 - \alpha u_3
	\end{array}\right].
\end{equation}
Since this multiplication is the local reaction component of the macroscopic model, we have
\begin{equation}
	\mathbf{G}(\mathbf{u}) = \left[\begin{array}{c}
	g_1(u_1,u_2,u_3) \\
	g_2(u_1,u_2) -\alpha u_2 + \alpha u_3\\
	g_2(u_1,u_3)+\alpha u_2 - \alpha u_3
	\end{array}\right],
\end{equation}
where
\[
	g_1(u_1,u_2,u_3) = \beta_1 - \mu_1 u_1 + \frac{\rho_1 u_1^2}{u_2+u_3}, \quad g_2(u_1,u_2) = \frac{\beta_2}{2} - \mu_2 u_2 + \frac{\rho_2 u_1^2}{2}.
\]
Thus the Jacobian of $\mathbf{G}$ at $\mathbf{u}^*$ becomes
\begin{equation}
	\nabla \mathbf{G}(\mathbf{u}^*) = \left[ \begin{array}{c c c}
		g_{1,1} & g_{1,2} & g_{1,3} \\
		g_{2,1} & g_{2,2} -\alpha, & g_{2,3} + \alpha \\
		g_{2,1} & g_{2,3} + \alpha & g_{2,2} - \alpha
	\end{array}\right],
\end{equation}
where the partial derivatives $g_{1,m} = \left.\partial g_1/\partial u_m\right|_*$ are
\begin{equation}
	g_{1,1} = -\mu_1 + \frac{2\rho_1 u_1^*}{u_2^* + u_3^*}, \quad g_{1,2} = -\rho_1 \left( \frac{u_1^*}{u_2^* + u_3^*} \right)^2, \quad g_{1,3}  = g_{1,2},
\end{equation}
and $g_{2,m} = \left.\partial g_2/\partial u_m\right|_*$
\begin{equation}
	g_{2,1} = \rho_2 u_1^*, \quad g_{2,2} = -\mu_2, \quad g_{2,3}  = 0.
\end{equation}
In particular, if $\beta_1 = \beta_2 = 0$, then the partial derivatives reduce to
\begin{equation} \label{b0}
	g_{1,1} = \mu_1, \quad g_{1,2} = - \frac{\mu_1^2}{\rho_1}, \quad g_{2,1} = \frac{\mu_2}{\mu_1} \rho_1, \quad g_{2,2} = -\mu_2.
\end{equation}

\subsection{Characteristic equation for the linearized macroscopic equations} \label{apdx:char}
The linear operator $\mathcal{L}(k)$ in the characteristic equation \eqref{chareq} can be reduced to the form
\begin{equation}
	\mathcal{L}(k) = \left[\begin{array}{c c c}
	2\kappa_1 q & 0 & 0 \\
	0 & \kappa_2(q - i \sqrt{1-(1+q)^2}) & 0 \\
	0 & 0 & \kappa_2(q + i \sqrt{1-(1+q)^2})
	\end{array}\right] + \nabla \mathbf{G}(\mathbf{u}^*)
\end{equation}
in accordance with the substitution $q = \cos(k) -1$.
The corresponding characteristic equation is 
\[
	p(q,\lambda) = -\lambda^3 + p_2(q) \lambda^2 + p_1(q) \lambda + p_0(q).
\]
where
\[
	p_j(q) = \sum_{m=0}^2 p_{j,m}q^m.
\]
The coefficients $p_{j,m}$ are given by
\begin{align}
	p_{2,0} &= g_{1,1} + 2g_{2,2} - 2\alpha, \nonumber\\
	p_{2,1} &= 2 (\kappa_1 + \kappa_2), \nonumber\\
	p_{2,2} &= 0,
\end{align}
\begin{align} \label{p1}
	p_{1,0} &= 2\alpha( g_{1,1} + g_{2,2}) -g_{2,2}^2 -2(g_{1,1}g_{2,2} - g_{1,2} g_{2,1}), \nonumber\\
	p_{1,1} &= 2\left[(\alpha- g_{2,2})(2\kappa_1 + \kappa_2) -g_{1,1}\kappa_2 + \kappa_2^2\right], \nonumber\\
	p_{1,2} &= - 4 \kappa_1 \kappa_2,
\end{align}
and
\begin{align}
	p_{0,0} &= (2\alpha - g_{2,2})(2g_{1,2}g_{2,1} - g_{1,1}g_{2,2}), \nonumber\\
	p_{0,1} &= -2 \left[ (2\alpha - g_{2,2})g_{2,2} \kappa_1 +(\alpha g_{1,1} - g_{1,1}g_{2,2} + g_{1,2}g_{2,1})\kappa_2 + g_{1,1} \kappa_2^2\right],\nonumber \\
	p_{0,2} &= -4\kappa_1 \kappa_2(\alpha -g_{2,2}+\kappa_2).
\end{align}
Note that if $\beta_1 = \beta_2 = 0$, then the coefficients $p_j(q)$ are independent of $\rho_1$ and $\rho_2$.
Moreover, equation \eqref{b0} implies that the partial derivatives are independent of $\rho_2$, and only $g_{1,2}$ and $g_{2,1}$ depend on $\rho_1$.
However, the latter coefficients always appear in $p_{j,m}$ as the product $g_{1,2}g_{2,1} = -\mu_1\mu_2$, which is independent of $\rho_1$.
This establishes the above claim.

The coefficient $p_j(q)$ also determines the Turing-Hopf bifurcation condition \eqref{THb} 
\[
	p_\text{TH}(q) = p_0(q) + p_1(q)p_2(q) = \sum_{m=0}^3 p_{\text{TH},m} q^m.
\]
In order to investigate its behavior over $q \in [-2,0]$, we first compute
\begin{align}
	p_{\text{TH},0} &= p_{0,0} + p_{1,0}p_{2,0} = -2 (g_{1,1} + g_{2,2}) \\
	&\quad \left[ 2\alpha^2 - \alpha(g_{1,1}+g_{2,2}) + (g_{1,1} g_{2,2} - g_{1,2} g_{2,1}) + g_{2,2}^2 - 2 \alpha g_{2,2}\right]. \nonumber
\end{align}
Using the fact that $g_{2,2} = - \mu_2$ and the stability condition \eqref{stabineq} 
\[
	g_{1,1} + g_{2,2} <0, \quad g_{1,1}g_{2,2} -g_{1,2}g_{2,1} >0,
\]
one can show $p_{\text{TH},0}> 0$. 
Similarly, we find the sign of higher oder coefficients. The leading coefficient is 
\begin{equation}
	p_{\text{TH},3} = -8\kappa_1 \kappa_2 (\kappa_1 + \kappa_2),
\end{equation}
which is negative. The next leading coefficient can be written as  
\begin{align}
	p_{\text{TH},2}&= 4\left[\kappa_2^3 - (g_{1,1}+g_{2,2})\kappa_2 (2\kappa_1 + \kappa_2) \right . \nonumber \\
	&\quad + \left . \alpha ( 2\kappa_1^2 + 4\kappa_1\kappa_2 + \kappa_2^2) + 2 \mu_2 \kappa_1 (1 + \kappa_2)\right],
\end{align}
which is positive. 
Introducing variables
\[
	\varphi_1 = - g_{1,1} + \mu_2, \quad \varphi_2 = g_{1,1}g_{2,2} - g_{1,2} g_{2,1}
\]
which are all positive, the first order coefficient takes the form of
\begin{align}
	- \frac{p_{\text{TH},1}}{2} &= 4\alpha^2(2\kappa_1 + \kappa_2) + 2\kappa_2^2 \mu_2 + \alpha [ 2\kappa_2^2 + 6 \kappa_1 ( \mu_2 + \varphi_1) + \kappa_2 ( 3 \mu_2 + 5 \varphi_1)] \nonumber \\
	&\quad + 2 \kappa_1 (\mu_2^2 + \mu_2 \varphi_1 + \varphi_2 ) + \kappa_2 ( \mu_2^2 + \mu_2 \varphi_1 + \varphi_1^2 + \varphi_2),
\end{align}
which right-hand side becomes positive. Thus, we have $p_{\text{TH},1}<0$.
From the above results, we conclude that $p_\text{TH}(q) >0$ for $q \in [-2,0]$.

\subsection{Components of correlation matrices in the mesoscopic model} \label{apdx:meso}
The non-zero components of $\mathbf{B}_{1,n}(\mathbf{u})$ are
\begin{align}
	[\mathbf{B}_{1,n}]_{1,1} &= \beta_1 + \mu_1 + \rho_1 \frac{u_{1,n}^2}{u_{2,n}+u_{3,n}}, \nonumber\\
	[\mathbf{B}_{1,n}]_{2,2} &= \alpha (u_{2,n} + u_{3,n}) + \beta_2 + \mu_2 u_{2,n} + \rho_2 u_{1,n}^2, \nonumber\\
	[\mathbf{B}_{1,n}]_{2,3} &= B_{32}^{1,n} = - \alpha (u_{2,n} + u_{3,n}), \nonumber\\
	[\mathbf{B}_{1,n}]_{3,3} &= \alpha (u_{2,n} + u_{3,n}) + \beta_2 + \mu_2 u_{3,n} + \rho_2 u_{1,n}^2, \nonumber\\
	[\mathbf{B}_{1,n}]_{1,2} &= B_{21}^{1,n} = B_{13}^{1,n} = B_{31}^{1,n} =0,
\end{align}
Similarly, the elements of the diagonal matrices $\mathbf{B}_{2,n}(\mathbf{u})$ and $\mathbf{B}_{3,n}(\mathbf{u})$ are
\begin{align}
	[\mathbf{B}_{2,n}]_{1,1} &= \kappa_1 (u_{1,n-1} + 2 u_{1,n} + u_{1,n+1}), \nonumber\\
	[\mathbf{B}_{2,n}]_{2,2} &= \kappa_2 (u_{2,n-1} + u_{2,n}), \nonumber\\
	[\mathbf{B}_{2,n}]_{3,3} &= \kappa_2 (u_{3,n} + u_{3,n+1}),
\end{align}
and
\begin{align}
	[\mathbf{B}_{3,n}]_{11} &= -\kappa_1 (u_{1,n-1} + u_{1,n}), \nonumber\\
	[\mathbf{B}_{3,n}]_{22} &= -\kappa_2 u_{2,n-1}, \nonumber\\
	[\mathbf{B}_{3,n}]_{33} &= -\kappa_2 u_{3,n}.\end{align}

\subsection{Coefficients in stochastic Turing pattern condition} \label{apdx:psd}
The first element of the power spectral density is given by equation \eqref{psdw}, and the coefficients have the following explicit forms:
\begin{align}
	\theta_{1,1} &= 2[\mathbf{B}_3^*]_{1,1},\nonumber\\
	\theta_{1,0} &= [\mathbf{B}_1^*]_{1,1} + [\mathbf{B}_2^*]_{1,1} + 2[\mathbf{B}_3^*]_{1,1},
\end{align}
\begin{align}
	\theta_{2,1} &= - 2\alpha \kappa_2 + 2 g_{2,2} \kappa_2 - 2\kappa_2^2,\nonumber\\
	\theta_{2,0} &= g_{2,2}^2 - 2\alpha g_{2,2},
\end{align}
and
\begin{align}
	\theta_{3,2} &= -4\left(\alpha \kappa_1 \kappa_2 - g_{2,2} \kappa_1 \kappa_2 + \kappa_1 \kappa_2^2\right),\nonumber\\
	\theta_{3,1} &= -2\left(\alpha g_{1,1} \kappa_2 + 2 \alpha g_{2,2} \kappa_1 - g_{1,1} g_{2,2} \kappa_2 +g_{1,1} \kappa_2^2 + g_{1,2} g_{2,1} \kappa_2 - g_{2,2}^2 \kappa_1\right),\nonumber\\
	\theta_{3,0} &= -2 \alpha g_{1,1} g_{2,2} + 4 \alpha g_{1,2} g_{2,1} + g_{1,1} g_{2,2}^2 - 2 g_{1,2} g_{2,1} g_{2,2}.
\end{align}

Taking the limit $\alpha \to \infty$ in the stochastic Turing condition \eqref{scritc}, we have a quadratic equation \eqref{ascondst} and its coefficients are the following:
\begin{align}
	\zeta_2 &= 2 g_{2,2}^2 u_1^*, \nonumber\\
	\zeta_1 &= \mu_1(1+ u_1^*)g_{2,2}^2 + 2 g_{1,2} g_{2,1} \kappa_2 u_1^* + g_{1,1} g_{2,2}^2 u_1^* - 2 g_{1,2}g_{2,1}g_{2,2}u_1^*, \nonumber \\
	\zeta_0 &= \mu_1(1+u_1^*)g_{1,2}g_{2,1} \kappa_2.
\end{align}

%------------------------------------------------------------------------------------------------------------------------------------------------------

\end{document}